# A High Rate Tension Device for Characterizing Brain Tissue

Badar Rashid[1], Michel Destrade[2,1], Michael Gilchrist[1,3*]

[1]School of Mechanical and Materials Engineering, University College Dublin, Belfield, Dublin 4, Ireland

[2]School of Mathematics, Statistics and Applied Mathematics, National University of Ireland Galway, Galway, Ireland

[3]School of Human Kinetics, University of Ottawa, Ontario K1N 6N5 Canada

*Corresponding Author

Tel: + 353 1 716 1884/1991, + 353 91 49 2344  Fax: + 353 1 283 0534

Email: Badar.Rashid@ucdconnect.ie (B. Rashid), michael.gilchrist@ucd.ie (M.D. Gilchrist), michel.destrade@nuigalway.ie (M. Destrade)

**Abstract** The mechanical characterization of brain tissue at high loading velocities is vital for understanding and modeling Traumatic Brain Injury (TBI). The most severe form of TBI is *diffuse axonal injury* (DAI) which involves damage to individual nerve cells (*neurons*). DAI in animals and humans occurs at strains > 10% and strain rates > 10/s. The mechanical properties of brain tissues at these strains and strain rates are of particular significance, as they can be used in finite element human head models to accurately predict brain injuries under different impact conditions. Existing conventional tensile testing machines can only achieve maximum loading velocities of 500 mm/min, whereas the Kolsky bar apparatus is more suitable for strain rates > 100/s. In this study, a custom-designed *high rate tension device* is developed and calibrated to estimate the mechanical properties of brain tissue in tension at strain rates ≤ 90/s, while maintaining a uniform velocity. The range of strain can also be extended to 100% depending on the thickness of a sample. The same apparatus can be used to characterize the dynamic behavior of skin and other soft biological tissues by using appropriately sized load cells with a capacity of 10 N and above.

# 1 INTRODUCTION

Over the past three decades, several research groups have investigated the mechanical properties of brain tissue over a wide range of loading conditions in order to elucidate the mechanisms of Traumatic Brain Injury (TBI). During severe impact to the head, brain tissue experiences compression, tension and shear; however, limited tests have been performed to analyze the behavior of tissue in tension [1-3]. To gain a better understanding of TBI, several research groups have developed numerical models which contain detailed geometric descriptions of anatomical features of the human head, in order to investigate internal dynamic responses to multiple loading conditions [4-12]. However, the fidelity and predictive accuracy of these models is highly dependent on the accuracy of the material properties, suitable to model impact conditions.

*Concussion* is the most minor and the most common type of TBI, whereas *diffuse axonal injury* (DAI) is the most severe form of injury which involves damage to individual nerve cells (*neurons*) and loss of connections among neurons. The DAI in animals and human has been estimated to occur at macroscopic shear strains of 10% – 50% and strain rates of approximately 10 – 50/s [13, 14]. Existing universal tensile machines have cross head speeds limited to 500 mm/min and in some cases to 2500 mm/min. It is therefore not possible to test soft biological tissues at a strain rate range of 10 – 50/s. The other available machinery is the Kolsky test apparatus, although it is more suitable for strain rates > 100/s. Recently, Tamura et al., [1] designed an apparatus to perform tests at 0.9, 4.3 and 25/s, although it is only the fastest of these rates that is close to real-world impact speeds.

In this study, a custom-designed *high rate tension device* (HRTD) is described which is capable of testing brain tissue up to a maximum strain rate of 90/s, at a uniform velocity. The maximum loading rate of this device approximately covers the entire range of strain rates as observed by various research groups during axonal injury investigations [13-20].

# 2 MATERIALS AND METHOD

## 2.1 Design specification

An apparatus was required to perform tensile tests on brain tissue at a strain rate range of 10 – 90/s and strain range of 10 – 100%. The range of strain and strain rates are based on the investigations conducted by various research groups [13-20]. The device should be capable of measuring reaction force (N) and displacement (mm) signals directly from the tissue during the extension phase at a uniform velocity. The system should have the capability to perform tensile tests at variable speeds from 100 to 1500 mm/s with high precision. The sample thickness should be selected for the tests so that results are not affected by stress wave propagation generated at a maximum strain rate of 90/s. Besides these requirements, the test protocol should provide sufficient information to prepare and mount tissue samples in a repeatable manner.

## 2.2 Construction and instrumentation

In order to perform tests at high loading velocities, programmable electronic actuators are usually used for testing soft biological tissues in compression or tension. All current actuators are designed specifically to produce successively acceleration, then uniform velocity, and then deceleration during the last phase of the travel. The deceleration phase before the end of the stroke poses a formidable challenge, when testing is required at higher strain rates > 10/s while simultaneously ensuring uniform velocity. The problem is further compounded when the amount of tissue extension during tensile tests is in the order of a few millimeters. All these factors were specifically addressed during the development process.

The HRTD is divided into a specimen testing mechanism and a striking mechanism based on its basic functioning, as shown in Fig. 1. The major components of the apparatus include a *servo motor controlled - LEFB32T-700 programmable electronic actuator* with a stroke length of 700 mm and a maximum velocity of 1500 mm/s, two *5 N load cells* (Transducer Techniques) with a rated output of 1.46 mV/V nominal and a *Linear Variable Displacement Transducer* (LVDT). The type ACT1000A LVDT developed by RDP electronics had a sensitivity of 16 mV/mm (obtained through calibration), range ± 25 mm, linearity ± 0.25, spring force at zero position 2.0 N and spring rate of 0.3 N/cm. The mechanism provides high repeatability in the positioning accuracy up to ± 0.1 mm.

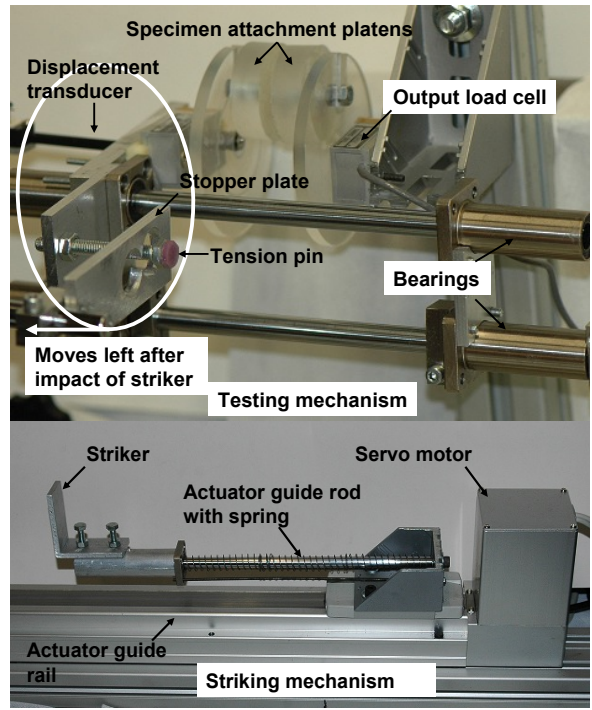

Fig. 1 The High Strain Rate Tension Device (HSTD), capable of testing brain tissue at high strain rate (≤ 90/s). The force (N) and displacement (mm) signals are received simultaneously through a data acquisition system (four channel Handyscope HS-4).

The testing mechanism is used to mount a cylindrical brain specimen between two platens in order to measure force and displacement. The striking mechanism is mainly composed of the *striker* with the guide rod, which is driven by the *servo motor,* and which impacts on the *tension pin*. The *tension pin* moves in a leftward direction which simultaneously moves the load cell to the left, thus generating tension in the brain tissue specimen. The displacement of the *tension pin* is controlled by the *stopper plate*. The force (N) sensed by the stationary load cell (output load cell) is used for further analysis. Load cells - GSO series -5 to +5 N (Transducer Techniques) were used for the experimentation. The rated output was 1.46 mV/V nominal with a safe overload of 150% of rated output. The excitation voltage applied to the load cell was 2.48 V DC and the amplified signal (amplification -101) was analyzed through a data acquisition system (4 channel Handyscope developed by TiePie company) with a sampling frequency of 10 kHz. Finally, the measured voltage signal (output) was converted to force (N) using the multiplication factor of 13.66 N/V[a] for further analysis.

## 2.3   Calibration to achieve uniform velocity

Calibration of the HRTD was essential in order to ensure uniform velocity during extension of brain tissue at each strain rate. Two main contributing factors for the non-uniform velocity were the deceleration of the electronic actuator when it is approaching the end of the stroke, and the opposing forces acting against the striking mechanism. Therefore, to overcome the deceleration of the electronic actuator, the striking mechanism (see Fig. 1) was designed to ensure that it impacts on the tension pin approximately 150 mm before the actuator comes to a complete stop. The *striker* impact generates backward thrust, which is fully absorbed by the spring mounted on the actuator guide rod in order to prevent any damage to the *programmable servo motor*.

---

[a] The corrected value of measured voltage = 2.48 (V) x1.46 mV/V x101 = 0.366 V. Thus the multiplication factor to convert measured voltage from 5N load cell becomes (5 N / 0.366 V) = 13.66 N/V.

The second important factor was the opposing forces acting against the striking mechanism. The LVDT probe inherently exerts 2.0 N force against the movement of the striker; moreover, the sliding components of the *testing mechanism* also provide resistance to any change in motion. Therefore, to achieve uniform velocity, the actual actuator velocity must be higher than the required (theoretically calculated) velocity to overcome these opposing forces. During the calibration process, the actuator was run several times to achieve uniform velocity. Fig 2 shows a typical output from an LVTD depicting displacement (mm) against time (ms) at a strain rate of 30/s, which shows that the accurate uniform velocity was successfully achieved. A similar procedure was adopted for all strain rates ≤ 90/s. Once the system is calibrated for a particular velocity, the displacement transducer and all other components must not be disassembled or changed.

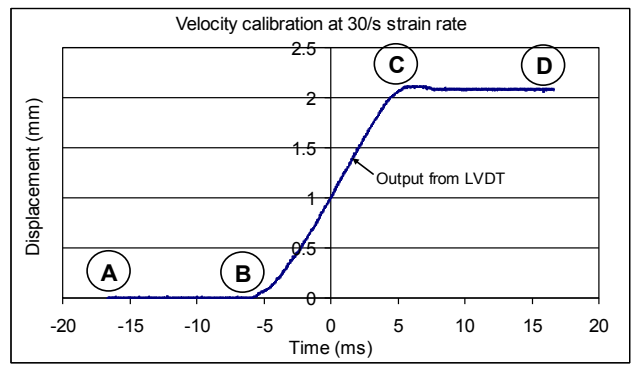

Fig. 2 A typical displacement (mm) – time (ms) output from an LVDT when a 7.0 mm thick cylindrical sample is extended up to 30% strain at a strain rate of 30/s.

Point A on Fig 2 depicts that the transducer is not moving but that it can still sense the signal. Displacement measurement starts at the moment the *striker* impacts on the *tension pin,* as depicted at point B. The displacement of the LVDT stops at point C and the signal continues to be acquired up to point D. The displacement between B to C actually corresponds to extension of brain tissue during tensile tests at high strain rates. The linear displacement profile between stages B and C shows that uniform velocity has been achieved between -5 ms < time < 5 ms.

## 2.4 Experimental protocol

***Specimen preparation****.* Ten fresh porcine brains from approximately six month old pigs were collected approximately 12 h after death from a local slaughter house. Porcine brains were preserved in a saline solution at 4 to 5 ºC during transportation, which took forty minutes. One half of the cerebral hemisphere of porcine brain was cut in the coronal plane and then cylindrical samples containing mixed white and gray matter were extracted from different regions of the brain, as shown in Fig. 3 (a). Cylindrical samples of nominal diameter 15.0±0.1 mm were cut using a circular steel die cutter as shown in Fig. 3. Variable thicknesses of cylindrical samples were obtained by inserting samples into cylindrical metal disks of different thicknesses (3.0 to 8.0 mm). The excessive brain portion was then removed with a surgical scalpel blade. Specimens were not all excised simultaneously, rather each specimen was tested first and then another specimen was extracted from the cerebral hemisphere. Each specimen was tested only once. This procedure was important to prevent the tissue from losing its stiffness and preventing dehydration (because of the viscoelastic nature of tissue) and thus contributed towards repeatability in the experimentation. Experiments were completed within 4 – 5 h post-mortem at a nominal room temperature of 22 ºC.

***Specimen mounting procedure.*** Dynamic tests (strain rate: 90/s) on HRTD and quasistatic tests (strain rate: 2/s) on a standard Tinius Olsen material testing machine (maximum speed limit: 500 mm/min) were performed on porcine brain tissue. Here, the reliable attachment of soft tissue to the platens for both the tests is very important in order to achieve high repeatability. For tests on HRTD, the surfaces of the platens were first covered with a masking tape substrate to which a thin layer of surgical glue (Cyanoacrylate, Low-viscosity Z105880–1EA, Sigma-Aldrich, Wicklow, Ireland) was applied. The prepared cylindrical specimen of brain tissue is placed between the platens. The two platens are separated by precisely machined spacers of variable thickness which corresponds to respective sample thicknesses (3.0 to 8.0 mm nominal); these ensure that the specimen is not overstressed as shown in Fig. 3(b). The relative displacement of platens is prevented in all directions by applying two clamps opposite to each other. Thereafter, the complete assembly is mounted on the testing mechanism and the platens are attached with the two load cells (fixed and movable). Lastly, the spacers between the platens are removed in such a manner that they do not touch the brain specimen. Approximately 3 – 4 minutes of settling time is given to ensure proper adhesion

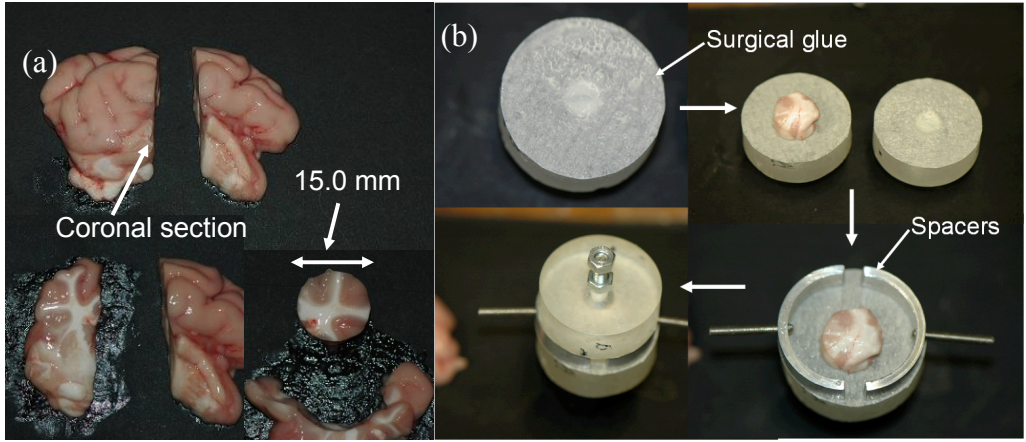

Fig. 3 – Experimental protocol (a) indicates extraction of cylindrical specimen from coronal plane of nominal diameter 15.0±0.1 mm (b) attachment procedure of specimen to the platens with the spacer. Spacers of variable thickness can be used corresponding to the thickness of the specimen.

This procedure facilitates excellent attachment of tissue to the platens and also serves to ensure no-slip boundary conditions. The distance between the platens is measured with a Vernier Caliper. Calibrating metal disks of variable thickness (3.0 to 8.0 mm) are also used to confirm the required distance between the platens before the start of experimentation.

## 3 RESULTS AND VALIDATION

### 3.1 Selection of sample thickness

Stress wave propagation effects were analyzed by taking variable sample thicknesses of porcine brain tissue according to the experimental protocol discussed above. The main purpose is to select a sample thickness which is least affected by stress wave propagation at a maximum strain rate of 90/s. Therefore, ten tests were conducted at each nominal sample thickness of 3.0 – 8.0 mm while maintaining a constant nominal diameter of 15.0 mm. Force (N) and displacement (mm) data were measured directly against time (s) through the data acquisition system. The data was then converted to engineering stress (kPa) and engineering strain. The engineering stress can be calculated from the force data by dividing it by the original cross-sectional area of the specimen. The nominal strain is obtained by dividing the measured displacement from the transducer by the initial cylindrical sample thickness. The results have been analyzed up to 30% strain at a maximum strain rate of 90/s, as shown on Fig 4.

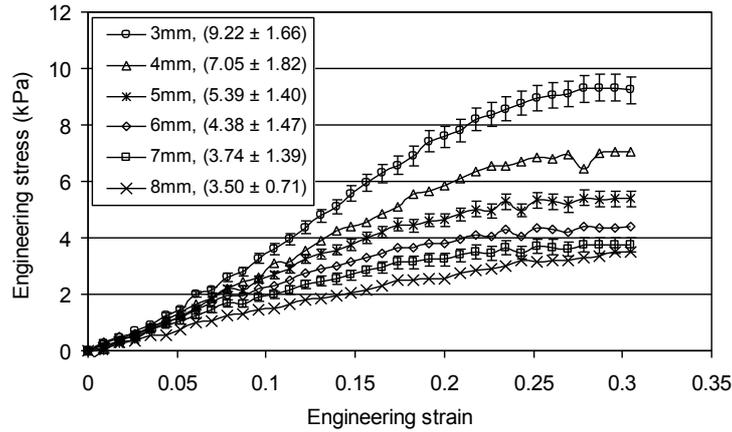

Fig. 4 – Tensile tests at different sample thicknesses (3.0 – 8.0 mm). Results for each thickness are the average of ten separate tests and indicated as mean ± SD.

It is quite evident that the stiffness is maximum at a sample thickness of 3.0 mm and minimum at 8.0 mm. The variation of stresses with sample thickness indicates that stress wave propagation effects and inhomogeneous deformation effects of brain tissue are dominant at lower sample thicknesses. Apparent elastic moduli, $E_1$, $E_2$ and $E_3$ were also calculated from the mean stress – strain curves (Fig. 4) corresponding to the strain ranges of 0 – 0.1, 0.1 – 0.2 and 0.2 – 0.3 respectively for each sample thickness as shown in Table 1.

Table 1. Apparent elastic moduli at each sample thickness.

| Young's modulus | Cylindrical sample thickness (mm) | | | | | |
|---|---|---|---|---|---|---|
| | 3.0 | 4.0 | 5.0 | 6.0 | 7.0 | 8.0 |
| $E_1$ (kPa) | 34.49 | 31.08 | 26.78 | 23.24 | 19.85 | 15.14 |
| $E_2$ (kPa) | 39.76 | 27.20 | 19.50 | 14.77 | 12.61 | 10.60 |
| $E_3$ (kPa) | 16.56 | 12.18 | 7.64 | 5.79 | 4.94 | 4.64 |

It is observed that the moduli, $E_1$, $E_2$ and $E_3$ are maximum at the smallest sample thickness (3.0 mm) and minimum at the largest thickness (8.0 mm). Moreover the moduli are significantly different at each sample thickness. At a sample thickness of 3.0 mm, the moduli, $E_1$, $E_2$ and $E_3$ are 56%, 75% and 72% higher than the sample thickness of 8.0 mm. However, the moduli, $E_1$, $E_2$, $E_3$ at a sample thickness of 7.0 mm are 24%, 16%, and 6% higher than the 8.0 mm sample thickness which are comparatively much less as compared to other thickness values. The sample thickness plays a critical role in the accurate estimation of stress values, particularly at higher strain rates as shown in Fig 4. Based on this analysis, cylindrical sample thicknesses of porcine brain tissue of 8.0mm or larger need to be used on the HRTD at the maximum strain rate of 90/s in order to avoid stress wave propagation effects and inhomogeneous deformation of the tissue. These results are based on samples of porcine brain tissue only; a similar procedure would need to be adopted to determine accurate sample thicknesses of other soft biological tissues with different geometries.

## 3.2 Validation of tensile test results

The experimental data obtained from the HRTD was further validated against the standard Tinius Olsen material testing machine as shown in Fig 5. The maximum speed limit of this machine was 500 mm/min (8.3333 mm/s), therefore testing was only possible at low strain rates < 10/s. A sample of nominal thickness 4.0 mm and diameter 15.0 mm was selected for testing on both machines (Tinius Olsen and HRTD). The specimen attachment procedure in the case of the Tinius Olsen machine was similar to the procedure discussed for the HRTD in Section 2.4. The surfaces of the top and lower platens were first covered with a masking tape substrate to which a thin layer of surgical glue was applied. The prepared cylindrical specimen of tissue was then placed on the lower platen. The top platen, which was attached to the 10 N load cell on the test machine, was than lowered slowly so as to just touch the top surface of the specimen. Four minutes settling time was sufficient to ensure proper adhesion of the specimen to the top and lower platens. The cylindrical brain specimens were then stretched to 50% strain at a velocity of 8.0 mm/s, which corresponded to a strain rate of 2/s. Fig. 5 (b) shows good agreement of experimental data between the Tinius Olson and HRTD testing machines. One of the major contributing factors to this good agreement is the low or quasi-static velocity (2/s) during these tensile tests, thus avoiding any error due to stress wave propagation. Results of tensile experiments conducted by other researchers [1-3] at quasi static loading conditions were superimposed on the results presented in this study as shown in Fig. 5 (b). Miller and Chinzei [2] performed tensile tests using cylindrical specimens (30.0 mm diameter and 10.0 mm height nominal) at strain rates of 0.64/s and 0.00064/s, similarly Tamura et al., [1] also conducted tests using cylindrical specimens (14.0 mm diameter and 14.0 mm height nominal) at strain rates of 25, 4.3 and 0.9/s. Velardi et al., [3] used rectangular specimens (nominal dimensions: 2.5 mm thick, 10.0 mm wide and 40 to 60 mm long) to perform tensile tests at a strain rate of 0.01/s. Only strain rates of existing studies [1-3] were selected for the comparison purpose which were close to a strain rate of 2/s (present study) as shown in Fig. 5 (b). Large variations in the experimental data is observed because of different test protocols, specimen geometry and loading conditions. However, experimental results of the present study are approximately in the same order of magnitude as observed in the case of Miller and Chinzei [2] as depicted in Fig. 5 (b).

To the best of authors' knowledge, there is no experimental data available in order to validate tensile test results at a strain rate of 90/s. Therefore validation was performed against finite element simulations using ABAQUS 6.9/Explicit. Material parameters used for the numerical analysis were derived by fitting the one-term Ogden model [21] to the average experimental data at a nominal specimen thickness of 8.0 mm and 15.0 mm diameter, which was stretched up to 30% strain at a strain rate of 90/s as shown in Fig. 4. The brain tissue was considered to be an isotropic, homogeneous, incompressible hyperelastic material. The initial shear modulus, $\mu$ = 4400 Pa, stiffening parameter, $\alpha$ = 0.13 and tissue density, $\rho = 1040 \, kg/m^3$ were used as input material parameters for the numerical simulations. C3D8R elements (8-node linear brick, reduced integration with relax stiffness hourglass control) were used. One side of the cylindrical specimen was constrained in all directions, whereas the other side was allowed to move at a particular velocity. The time period for the simulation was adjusted to achieve required amount of strain in the brain specimen as shown in Fig.5 (c). The average surface force (N) was divided by the original cross sectional area ($m^2$) to estimate engineering stress (Pa). An excellent agreement is achieved between the numerical and experimental results as shown in Fig. 5 (d).

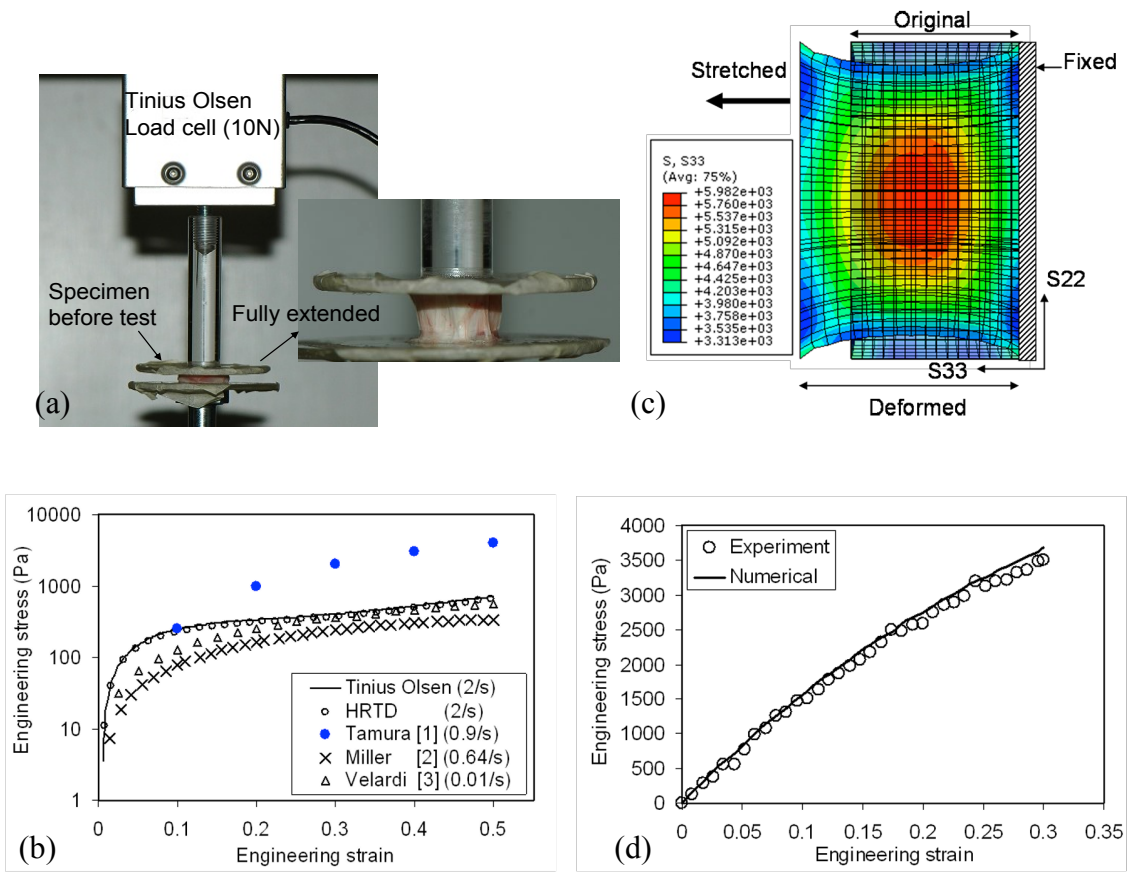

Fig. 5 – Validation of tensile tests carried out against standard test setup. (a) The velocity on Tinius Olsen machine was 8.0 mm/s corresponding to a strain rate of 2/s. (b) Good agreement of engineering stress is achieved between Tinius Olsen and HRTD test results. Results from previous studies (1-3) are also superimposed for comparison. (c) Numerical simulation using material parameters at 90/s strain rate of 8.0 mm thick specimen. (d) Excellent agreement is achieved between numerical engineering stress and experimental engineering stress at 90/s strain rate.

## 4 DISCUSSION

It is possible to determine the mechanical properties of porcine brain tissue at a high strain rate $\leq 90$/s at variable strains by using this HRTD. Reliable experimental data can be obtained from the device, by carefully selecting the sample thickness and performing calibration before the actual tests. The specimen preparation protocol and mounting procedure on the testing mechanism are crucial in achieving consistency and repeatability in the experiments.

The HRTD was specifically designed for the testing of porcine brain tissue; therefore it was more appropriate to use brain tissue instead of any other soft biological tissue for the calibration of the test rig and selection of most suitable sample thickness. However, brain tissue is among the most difficult of biological materials to handle because of its inherent sticky nature and because it degrades with time. Due to the non-availability of any other standard machine to perform tests at higher strain rates > 10/s, validation of the HRTD was carried out against the standard Tinius Olsen testing machine at a quasi-static velocity of 8 mm/s.

The HRTD is most suitable for testing cylindrical specimens under tension. However, simple shear tests at high strain rates can also be performed easily on this apparatus simply by replacing the specimen attachment platens as shown in Fig 6. The horizontal distance between the platens can be adjusted based on the dimension of test specimen and platens. Fig. 6 (b) shows a rectangular shaped specimen attached between the platens using surgical glue. The top platen remains fixed while the bottom platen moves to the left side, thus causing simple shear in the brain tissue at a constant strain rate. It is not possible to use this device for compression tests.

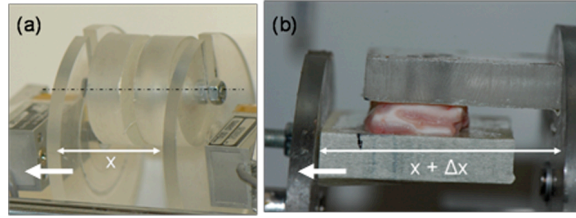

Fig. 6 – Tension tests (a) and shear tests (b) can be performed on HRTD by replacing the platens and adjustment of horizontal distance.

If the LVDT or any component is replaced or reinstalled, the device must be recalibrated for the required uniform velocity before undertaking any test. A temperature controlled chamber could be developed to completely surround the specimen testing mechanism in order to maintain particular temperatures, if required.

# 5 CONCLUSION

The mechanical characterization of soft biological tissues at high strain rates has always been a challenge for researchers. In this study, we have developed a HRTD which can be utilized effectively to extract force and displacement data for brain tissue at a high strain rate ($\leq$ 90/s). Simple shear tests at high strain rates can be performed on this device simply by replacing the specimen attachment platens.

**Acknowledgements**   The authors thank John Gahan, Tony Dennis and Pat McNally of University College Dublin for their assistance in machining components and developing electronic circuits for the experimental setup. This work was supported for the first author by a Postgraduate Research Scholarship awarded by the Irish Research Council for Science, Engineering and Technology (IRCSET), Ireland.